\documentclass[prl,aps,twocolumn,showpacs,preprintnumbers,amsmath,amssymb,
superscriptaddress,notitlepage]{revtex4}

\usepackage{graphicx}
\usepackage{dcolumn}
\usepackage{bm}
\usepackage{hyperref}

\font\msytw=msbm9 scaled\magstep1 
scaled\magstep1

\let\a=\alpha \let\b=\beta    \let\d=\delta \let\e=\varepsilon
  \let\h=\eta     
\let\m=\mu    \let\n=\nu         \let\p=\pi    \let\r=\rho
\let\s=\sigma    \let\f=\varphi 
   \let\o=\omega
   \let\L=\Lambda \let\X=\Xi
         
\let\O=\Omega 

 \def\VV{{\cal V}}
 \def\WW{{\cal W}}
 \def\BBB{{\cal B}}
\def\RR{{\cal R}}\def\LL{{\cal L}}  
 \def\SS{{\cal S}}

   \def\pp{{\bf p}}
 \def\xx{{\bf x}} \def\yy{{\bf y}} 

\def\kk{{\bf k}}

 \def\ZZZ{\hbox{\msytw Z}}

\def\\{\hfill\break}
\let\==\equiv
\let\io=\infty
\let\0=\noindent
\def\media#1{{\langle#1\rangle}}

\def\tende#1{\,\vtop{\ialign{##\crcr\rightarrowfill\crcr\noalign{\kern-1pt
    \nointerlineskip} \hskip3.pt${\scriptstyle #1}$\hskip3.pt\crcr}}\,}
\def\otto{\,{\kern-1.truept\leftarrow\kern-5.truept\to\kern-1.truept}\,}

\def\to{\rightarrow}

\def\qed{\hfill\raise1pt\hbox{\vrule height5pt width5pt depth0pt}}

\def\be{\begin{equation}}
\def\ee{\end{equation}}
\def\bea{\begin{eqnarray}}
\def\eea{\end{eqnarray}}
\def\nn{\nonumber}
\def\pref#1{(\ref{#1})}

\begin{document}

\title{Universal conductivity and dimensional crossover in multi-layer graphene}
\author{Vieri Mastropietro}%
\affiliation{%
Universit\`a di Roma Tor Vergata, Viale della Ricerca Scientifica
00133 Roma, Italy,EU}

\begin{abstract}
We show, by exact Renormalization Group methods, that in
multi-layer graphene the dimensional crossover energy scale is
decreased by the intra-layer interaction, and that for
temperatures and frequencies greater than such scale the
conductivity is close to the one of a stack of independent layers
up to small corrections.
\end{abstract}
\pacs{71.10.Fd, 72.80.Vp, 05.10.Cc, 05.30.Fk} \maketitle

Recent experiments \cite{N1} have found that the optical
conductivity in multi-layer graphene is essentially constant in a
wide range of frequencies and equal to $N \s_0$, with $N\le 5$ the
number of layers and $\s_0={\pi e^2\over 2 h}$; remarkably, a
value both independent from the intra-layer hopping $t\sim 3eV$
and inter-layer hopping $t^\perp\sim 0.3eV$ and depending only on
the conductivity quantum $h/e^2$ and the number of layers $N$. In
the absence of interaction, when $N=1$ (monolayer graphene) this
observation finds a nice explanation in the effective description
of fermions on the honeycomb lattice in terms of massless Dirac
fermions in $2+1$ dimensions, whose conductivity in the limit of
zero frequency is exactly $\s_0$, as was shown in \cite{L}. The
band structure of multi-layer graphene is completely different and
the description in terms of Dirac fermions breaks down at low
energies; in the case of bilayer graphene, for instance,
\cite{MF}, a quadratic spectrum instead of a linear relativistic
one is found for certain kind of hopping terms. However as pointed
out in \cite{N1,SPG} in multi-layer graphene there is a
dimensional crossover scale $t_\perp$ and at energy scales greater
than $t_{\perp}$ the conductivity of multi-layer graphene is equal
to the one of a stack of $N$ independent layers up to small
corrections.

The above reasoning neglects the role of the interactions among
charge carriers, which are well known to be rather strong and
producing observed important effects in several physical
observables \cite{E}. In the case of monolayer graphene there are
several papers discussing the role of interactions on the
conductivity \cite{M2,SS2,H1,K,GMP1}, but in the case of
multi-layer this problem is much less studied, even if
interactions in multi-layer graphene are believed to radically
alter the low energy properties and spontaneous symmetry breaking
is expected \cite{V5,V6}.

In the present paper we will consider a system of several layers
in which the electrons can hop from a plane to another with
coupling $t^\perp$, and $t^\perp<<t$; each layer is described in
terms of interacting fermions on the honeycomb lattice with an
electromagnetic (e.m.) interaction \cite{GGV,GMP2}. We will show
that the interaction produces a renormalization of the dimensional
crossover scale and the conductivity of multi-layer graphene is
equal to the one of a stack of $N$ independent layers up to small
corrections at energy scales greater than
\be t^{*,\perp}\sim t^{\perp} ({t^{\perp}\over t})^{\h\over
1-\h}\ee
where $\h>0$ is the exponent of the wave function renormalization.
Note that the crossover scale is {\it decreased} by the
intra-layer interaction, a phenomenon strongly resembling what
happens in fermionic chains, see {\it e.g.} \cite{2}. Even if our
results are found for values of the parameters close to the
infrared fixed point of monolayer graphene (that is, the Fermi
velocity large enough), they provide an evidence that the
many-body interaction preserves and even enforces the conductivity
universal behavior, in qualitative agreement with experiments
\cite{N1}.

We consider the Hamiltonian for multi-layer graphene
\be H=\sum_{\a=1}^N H_\a+H_\perp \ee
where $H_\a$ describe a single graphene layer; more exactly
fermions hopping on the honeycomb lattice interacting through an
e.m. field introduced via the Peierls substitution \cite{GMP2}
\bea
&&H_\a=-t\sum_{\substack{\vec x\in\L_A \\
j=1,2,3}}a^{+}_{\vec x,\a} b^{-}_{\vec x + \vec \d_j,\a}
e^{ie\int_0^1\vec\d_j\cdot\vec A_\a(\vec x+s\vec\d_j,0)\,ds}
 + c. c.+\nn\\
&&\frac{e^2}2\sum_{\vec x,\vec y\in\L_A\cup\L_B} (n_{\vec
x}-1)\f_\a({\vec x}-{\vec y})(n_{\vec y}-1) \eea
$\psi_{\vec x,\a}^\pm=(a^\pm_{\vec x,\a}, b^\pm_{\vec x + \vec
\d_1,\a})= |\BBB|^{-1}\int_{\vec k\in\BBB}d\vec k\,\psi^\pm_{\vec
k,i,\a} e^{\pm i\vec k\vec x}$ for electrons with plane index
$\a=1,2$ and sitting at the sites of the two triangular
sublattices $\L_A$ and $\L_B$ of a honeycomb lattice.  We assume
that $\L_A$ has basis vectors $\vec l_{1,2}= \frac12(3,\pm\sqrt3)$
and that $\L_B=\L_A+\vec\d_j$, with $\vec \d_1=(1,0)$ and
$\vec\d_{2,3}=\frac12 (-1,\pm\sqrt3)$ the nearest neighbor
vectors; $\BBB$ is the first Brillouin zone and $\vec A_\a$ and
$\phi_\a$ are the respectively the vector field and the coulomb
potential on the plane $\a$.

Finally $H_{\perp}$ describes the hopping between graphene layers
and is assumed of the form \be
H_{\perp}=t^\perp\sum_{\a=1}^{N-1}\sum_{\vec x} j^{\perp}_{\a,\vec
x}\ee where $j^{\perp}_{\a,\vec x}$ is a bilinear operator
describing the hopping from layer $\a$ to layer $\a+1$, see {\it
e.g.} \cite{MF}.
%
%
The {\it planar paramagnetic current} is defined as
\be\vec J_{\vec p}=iet\sum_{\a=i}^{N}\sum_{\vec
x\in\L_A,j}\,e^{-i\vec p\vec x} \vec \d_j\h^j_{\vec
p}\big(a^+_{\vec x,\a}b^-_{\vec x+\vec \d_j,\a}- b^+_{\vec x+\vec
\d_j,\a} a^-_{\vec x,\a}\big)\nn \ee
where $\h^j_{\vec p}=\frac{1-e^{-i\vec p\vec \d_j}}{i\vec p\vec
\d_j}$. The two components of the paramagnetic current $\vec
J_{\vec p}$ will be seen as the spatial components of a
``space-time" three-components vector $\hat J_{\vec p,\m}$,
$\m=0,1,2$, with $\hat J_{\vec p,0}=e\hat\r_{\vec p}$ and
$\hat\r_{\vec p}$ the density operator.
%

If $O_{\xx}=e^{x_{0}H_\L}O_{\vec x_i}e^{-x_{0}H_\L}$, with
$\xx=(x_{0},\vec x)$, we denote by $\media{O^{(1)}_{\xx_1}\cdots
O^{(n)}_{\xx_n}}$ the thermodynamic limit of $\X^{-1} {\rm
Tr}\{e^{-\b H}{\bf T}(O^{(1)}_{\xx_1}\cdots O^{(n)}_{\xx_n})\}$,
where $\X={\rm Tr}\{e^{-\b H}\}$ and ${\bf T}$ is the operator of
fermionic time ordering. The current-current functions $\hat
K_{\m\n}(\pp)$ is defined as the 2D Fourier transforms of
$\media{J_{\xx,\m};J_{\yy,\n}}_{\b}$ and the {\it conductivity} is
\cite{SPG} (here $l,m=1,2$):
\be \s_{lm}(\o)= -\frac{2}{3\sqrt3}{1\over \o} [\hat
K_{l,m}(\o,0)-\hat K_{l,m}(0,0)]\label{rr} \ee
%
%
where $3\sqrt3/2$ is the area of the hexagonal cell of the
honeycomb lattice. In our notations, $\pp=(\o,\vec p)$, with
$\o\in \frac{2\p}{\b}\ZZZ$ the Matsubara frequency.


It is convenient to introduce the following {\it Grassman
integral}
\be e^{W(J)}=\int P(d\Psi)\int P(dA)e^{\VV(A+J,\psi) }\label{por}
\ee
where: $\psi^\pm_{\kk}$ are Grassman variables $(\kk=(k_0,\vec
k))$ and $P(d\psi)$ is the fermionic gaussian integration with
inverse propagator \be g^{-1}(\kk)= -\left(\begin{array}{cc} ik_0
& v \O^*(\vec k)
\\ v \O(\vec k) & ik_0 \end{array}\right)\;\label{vo}\ee with $v={3\over 2}t$
$\O(\vec k) = \frac23\sum_{j=1,2,3}e^{i\vec k(\vec \d_j -
\vec\d_1)}$ (note that $g(\kk)$ is singular only at the Fermi
points $\kk=\kk_F^{\pm}
=(0,\frac{2\p}{3},\pm\frac{2\p}{3\sqrt3})$); if $\m=0,1,2$,
$A^\m_\a(\pp)$ are gaussian variables with propagator $
\d_{\a,\b}\d_{\m\n}\frac{\chi(\pp)}{|\pp|}$ where $\chi$ act an an
ultraviolet cut-off; finally $\VV$ is the interaction whose
explicit form can be easily inherited from $H$. By suitable
derivatives with respect to $J$ the current-current correlation
can be obtained; note that we have exploited gauge invariance to
write the photon propagator in the Feynman gauge.
The generating function \pref{por} can be computed by exact
Renormalization Group methods. After the integration of the fields
$\psi^{(1)},A^{(1)},...,\psi^{(h+1)},A^{(h+1)}$ we get
\be e^{\WW(J)} =\int P(d\psi^{(\leq h)})P(dA^{(\leq h)})
e^{\VV^{(h)}(\sqrt{Z_h}\Psi^{(\leq h)},A^{(\leq
h)}+J)}\;,\label{3.6} \ee
where $P(d\psi^{(\le h)})$ is the fermionic integration with
propagator, if $r=\pm$ is the valley index
\bea&& \hat g^{(\leq h)}_{r}(\kk') =\nn\\
&& - \frac{\chi_h(\kk')}{Z_h}\begin{pmatrix} ik_0 & v_h\O^*(\vec
k'+ \vec p_F^{\,r})\\  v_h\O(\vec k'+ \vec p_F^{\,r}) & ik_0
\end{pmatrix}^{\!\!\!-1}\!\!\!\!\nn
\eea where $Z_h$ is effective wave renormalization and $v_h$ the
effective Fermi velocity, $P(dA^{(\le h)})$ is the gauge field
integration with propagator
$\d_{\a,\b}\d_{\m,\n}\frac{\chi_{h}(\pp)}{2|\pp|}$, with
$\chi_h(\kk'), \chi_h(\pp)$ smooth cut-off functions with support
smaller than $t 2^h$; moreover $\VV^{(h)}$ is the {\it effective
potential} which has the form
\be \VV^{(h)}(\psi,A)= \int d\underline \xx d\underline\yy
W^{(h)}_{n,m}\prod_{i=1}^n\psi^{\e_i}_{\xx_i,r_i,\a_i}\prod_{i=1}^m
A_{\m_i,\yy_i}\ee
where the kernels $W^{(h)}_{n,m}$ depends on the effective charge
at scale $h$ $e_{\m,h}$ and the effective hopping $t^\perp_h$. We
have now to describe the integration of the field $\psi^{(h)},
A^{(h)}$ and in this way we will iteratively define the effective
constants $Z_h,v_h,e_{\m,h},t^\perp_h$. In order to do that we
have to decompose $\VV^{(h)}$ as $\VV^{(h)}=\LL
\VV^{(h)}+\RR\VV^{(h)}$
with $\RR=1-\LL$; $\LL \VV^{(h)}$ is the {\it relevant} or {\it
marginal} part of the effective interaction while $\RR\VV^{(h)}$
is the {\it irrelevant} part. Generally this decomposition is
dictated by the naive scaling dimension which in the present case
is given by
\be D=3-n-m\label{dim} \ee
%
%
%
$\LL$ should select the terms with positive or vanishing dimension
$D$. However, if $h_\b$ is the temperature scale $\b t= 2^{-h_\b}$
and if the temperature verifies the condition
\be t 2^{h_\b}>t_{h_\b}^{\perp}\label{21} \ee
where $t_h^\perp$ is the hopping at scale $h$, there is an
improvement with respect to naive power counting, and certain
terms which are dimensionally relevant or marginal are indeed
irrelevant. In order to verify this fact, we can split the kernels
as
\be W_{n,m}^{(h)}=W_{n,m}^{(a)(h)}+W_{n,m}^{(b)(h)} \ee
where $W_{n,m}^{(a)(h)}$ is obtained from $W_{n,m}^{(h)}$ setting
$t^\perp=0$. Note that $W_{n,m}^{(a)(h)}$ in correspondence of
external fields with different plane index are vanishing.

We define the $\LL$ operator as \be\LL \hat
W^{(h)}_{2,1}(\kk')=\hat W^{(a)(h)}_{2,1}(0)\label{gggg}\ee
so that
\be \RR \hat W_{2,1}^{(h)}(\underline\kk')=[\hat
W_{2,1}^{(a)(h)}(\kk')-\hat W_{2,1}^{(a)(h)}(\underline{\bf
0})]+\hat W_{2,1}^{(b)(h)}(\kk')\label{hh} \ee
The first term in the r.h.s. of \pref{hh} can be rewritten as
$\underline\kk'\cdot\underline{\bf \partial}W_{n,m}^{(a)(h)}$, and
this produces an improvement $\sim 2^{h'-h}$ in the bound of the
kernel, if $h'$ is the scale of the momentum, which is sufficient
to make it irrelevant. Similarly the second term in \pref{hh},
namely $\hat W_{2,1}^{(b)(h)}(\hat\kk')$, has an extra $t^\perp_h
2^{-h} \le  2^{(h_\b-h)(1-\h)}$, $\h=O(e^2)$ (see below) with
respect to the bound for $W_{2,1}^{(h)}$, which again is enough to
make it irrelevant; therefore, the true marginal contribution is
given by the r.h.s. of \pref{gggg}. Regarding the terms quadratic
in the gauge fields, \be\LL \hat W^{(h)}_{0,2}(\pp)=\hat
W^{(a)(h)}_{0,2}(0)+\pp\partial \hat W^{(a)(h)}_{0,2}(0)\ee where
we have used that $\hat W^{(b)(h)}_{0,2}(0)$ has an extra
$(2^{-h}t^\perp_h)^2$ with respect to the naive dimension. Finally
the terms quadratic in the fermionic variables, if they have the
same plane index then \be\LL \hat W^{(h)}_{2,0}(\kk')=\hat
W^{(a)(h)}_{2,0}(0)+\kk'\partial \hat W^{(a)(h)}_{2,0}(\kk')\ee
where we have used that in $\hat W^{(b)(h)}_{2,0}$ there is an
extra gain $O((t^\perp_h 2^{-h})^2$, due to the conservation of
the plane index $\a$. On the other hand for the quadratic terms
with external fields corresponding to $j^\perp$ we define $\LL
\hat W^{(h)}_{2,0}(\kk')=\hat W^{(h)}_{2,0}({\bf 0})$, while for
the other terms with different layer index $\LL \hat
W^{(h)}_{2,0}(\kk')=0$.

The terms in $\partial_0 \hat W^{(a)(h)}_{2,0}$ and $\partial_1
\hat W^{(a)(h)}_{2,0}$ with both fields with the same plane index
are included in the free fermionic intergration and produces the
new effective wave function renormalization $Z_{h-1}$ and Fermi
velocity $v_{h-1}$; therefore we get after rescaling
\bea &&e^{\WW(J)} =\int P_{Z_{h-1},v_{h-1}}(d\psi^{(\leq
h-1)})P(dA^{(\leq
h-1)})\nn\\
&&e^{\tilde\VV^{(h)}(\sqrt{Z_{h-1}}\psi^{(\leq h)},A^{(\leq
h)}+J))}\;,\label{3.6} \eea
with
\bea &&\LL\tilde\VV^{(h)}(\psi^{(\leq h)}, A^{(\leq h)})=
\frac1{\b|\SS_L|}\sum_{\m,\pp}\,Z^{(\m)}_{h}e \hat
\jmath^{(\leq h)}_{\m,\pp}\hat A^{(\leq h)}_{\m,\pp}+\nn\\
&&\sum_\a 2^{h}\n_{\m,h}\hat A^{(\leq h)}_{\m,\a,-\pp}\hat
A^{(\leq h)}_{\m,\a,\pp}\big] +t^\perp_h\int d\xx
\sum_{\a=1}^{N-1} j^{(\le h)\perp}_{\xx,\a} \label{3.14b} \eea
where $\hat\jmath^{(\leq h)}_{\m,\pp}$ is the intra-layer current
and $j^{(\le h)\perp}_{\xx,\a}$ the inter-layer current. Note that
by construction the effective Fermi velocity $v_h$, effective wave
function renormalization $Z_h$ and effective charge $e
{Z^{\m}_h\over Z_h}\equiv e_{\m,h}$ are the same as in the
$t^{\perp}=0$ case and by exploiting Ward Identities we get, see
\cite{GMP2} $\n_h=O(e^2)$ and
\be {Z^{(0)}_h\over Z_h}=1+O(e)\quad\quad {Z^{(i)}_h\over Z_h
v_h}=1+O(e)\label{1a}\ee
Moreover, see \cite{GMP2}, the wave function renormalization
diverges with a power law with a critical exponent and the
effective Fermi velocity increases up to the light velocity with a
power law
\be Z_h\sim 2^{\h h}\quad\quad 1-v_h\sim 2^{\tilde h h}\ee
with $\h=\frac{e_{-\io}^2}{12\pi^2}+O(e_{-\io}^3)$ and $\tilde\h=
\frac{2e_{-\io}^2}{5\pi^2}+O(e_{-\io}^3)$.

Regarding the flow of $t^\perp_h$ we obtain
\be t^\perp_{h-1}={Z_h\over
Z_{h-1}}(t^\perp_h+\b_t^{(h)})\label{sol11} \ee
with $|\b_t^{(h)}|\le C_1 t^\perp_h e^6(t^\perp_h 2^{-h})^2$. It
is easy to see by induction that $|Z_h t^\perp_h-t^\perp|\le C_2
t^\perp e^6$. Assume indeed that it is true for $k\ge h$;
therefore
\be |t^\perp_{h-1} Z_{h-1}-t^\perp|\le  2 t^\perp C e^6
\sum_{k=h}^0(t^\perp_k 2^{-k})^2\ee
from which the inductive assumption follows. Note that the
effective hopping, even if {\it relevant} in the RG sense
according to naive power counting, remains small in this region of
temperatures. Moreover, from \pref{21} we obtain the condition
between the temperature and the hopping
\be \b^{-1}\ge t^{\perp} ({t_\perp\over t})^{\h\over 1-\h}\equiv
t^{*,\perp}\label{jjjj}\ee
As the flow of the effective parameters corresponding to the
relevant and marginal operators is bounded, the following bound is
obtained, for $h\ge h_\b$ (order by order in the renormalized
expansion) \be{1\over \L\b}\int d\underline\xx
|W^{(h)}_{n,m}(\underline\xx)|\le C 2^{h(3-n-m)}\label{rrr}\ee
We apply the above bound to the conductivity, which is given by
\be \s_{ii}(\o)= -\frac{2}{3\sqrt3}{1\over \o} \int dx_0 (e^{i\o
x_0}-1) K_{i,i}(\xx)\label{rr1} \ee
We can decompose
$K_{i,i}(\xx)=K^{(a)}_{i,i}(\xx)+K^{(b)}_{i,i}(\xx)$ where
$K^{(a)}_{1,1}(\xx)$ is obtained from $K_{i,i}(\xx)$ setting
$t^\perp=0$ and, for any $M$
\bea && |K^{(a)}_{i,i}(\xx)|\le [{Z_h^{(i)}\over Z_h}]^2
{2^{4h}\over
1+(2^h |\xx|)^M}\nn\\
&& |K^{(b)}_{i,i}(\xx)|\le [{Z_h^{(i)}\over Z_h}]^2 [{t_h\over
2^h}]^2 {2^{4h}\over 1+(2^h |\xx|)^M} \eea
The above estimates can be derived from the dimensional bound
\pref{rrr}; roughly speaking there is, with respect to \pref{rrr},
an extra $2^{3 h}$ due to a lacking integration and a decaying
factor ${1\over 1+(2^h |\xx|)^M}$ which can be extracted from the
chain of propagators connecting the external fields. In the bound
for $K^{(b)}_{i,i}(\xx)$ there are also two extra ${t_h\over
2^h}$. Therefore
\be \s_{ii}(\o)=\s_{ii}(\o)|_{t^\perp=0}+R^{(b)}_{ii}(\o)\ee
where
\be R^{(b)}_{ii}(\o)= -\frac{2}{3\sqrt3}{1\over \o} \int dx_0
(e^{i\o x_0}-1) K^{(b)}_{i,i}(\xx)\label{rr1}\ee
so that for $t^{*.\perp}\le \b^{-1}$
\bea &&|R^{(b)}_{ii}(\o)|\le C \sum_{h=h_\b}^1
(t^\perp)^2\int d\xx{2^{(2+2\h)h}\over 1+(2^h |\xx|)^5}\nn\\
&&\le {C\over\o}\sum_{h=h_\b}^1( t^\perp)^2 2^{h(-1+2\h)}\le C
({t^{*.\perp}\over\o})^{1-\h} (t^{*,\perp}\b)^{1-\h}\nn \eea
Therefore for $t^{*.\perp}\le \b^{-1}<< \o$ the conductivity is
the one of a stack of independent layers plus a negligible
corrections, in qualitative agreement with the observations in
\cite{N1}.

{\it Acknowledgements} The Author gratefully acknowledges
financial support from the ERC Starting Grant CoMBoS-239694

\end{document}